\documentclass[twocolumn,amsmath,amssymb,showpacs]{revtex4}
\usepackage{graphicx}

\begin{document}


\title{Kochen-Specker Sets with a Mixture of 16 Rank-1 and 14 Rank-2 Projectors for a Three-Qubit System}


\author{S.P.Toh\footnote{SingPoh.Toh@nottingham.edu.my; singpoh@gmail.com}}

\affiliation {Faculty of Engineering, The University of Nottingham Malaysia Campus,
Jalan Broga, 43500 Semenyih, Selangor Darul Ehsan, Malaysia.}


\date{\today}

\begin{abstract}
\emph{Kochen-Specker (KS) theorem denies the possibility for the noncontextual hidden variable theories to reproduce the predictions of quantum mechanics. A set of projection operators (projectors) and bases used to show the impossibility of noncontextual definite values assignment is named as the KS set. Since one KS set with a mixture of 16 rank-1 projectors and 14 rank-2 projectors proposed in 1995 [Kernaghan M and Peres A 1995 Phys.\ Lett.\ A \textbf{198} 1] for a three-qubit system, there are plenty of the same type KS sets and we propose a systematic way to produce them. We also propose a probabilistic state-dependent proof of the KS theorem that mainly focuses on the values assignment for all the rank-2 projectors.}
\end{abstract}

\pacs{03.65.Aa, 03.65.Ta, 42.50.Dv}

\maketitle


The Kochen-Specker (KS) theorem is another major theorem of impossibility of hidden variables in quantum mechanics (QM) besides the Bell theorem. While Bell theorem rules out the local hidden variable theories, KS theorem refutes the noncontextual hidden variable (NCHV) theories. The NCHV theories assume that the result of a measurement of an observable is predetermined and independent of the other mutually compatible or co-measurable measurements that may perform previously or simultaneously. QM is contextual, its measurement outcomes depend on the context of measurement.

To be precise, the KS theorem asserts that, in a Hilbert space of dimension $d \geq 3$, it is impossible to associate definite numerical values, 1 or 0, with every projection operator (projector) $P_i$, in such a way that, if a set of commuting $P_i$ satisfies $\sum P_i =I$, the corresponding value functions, namely $v(P_i)=0$ or 1, also satisfy $\sum v(P_i) =v(I)$ \cite{R1}. Kochen and Specker gave the first proof of KS theorem using a set of 117 projectors in three-dimensional real Hilbert space ($\Re^3$) \cite{R2}. The number of projectors reduced to 33 almost twenty-five years later \cite{R3}, and so far the most economical proof due to Conway and Kochen uses only 31 projectors \cite{R4}. There are also proofs of KS theorem for higher dimensional spaces, including $\Re^4$ \cite{R5, R6, R7}, $\Re^5$, $\Re^6$, $\Re^7$ \cite{R8},  $\Re^8$ \cite{R9, R10} and $\Re^{16}$ \cite{R11}.

Many experiments have been performed to show that quantum contextuality cannot be explained by any NCHV theories. The physical systems involved in the experiments include photons \cite{R12, R13, R14, R15}, neutrons \cite{R16, R17} and trapped ions \cite{R18}. Experiment is also proposed to test the KS theorem at a macroscopic level with superconducting quantum circuit \cite{R19}. Unlike the Bell theorem, the proof of the KS theorem requires neither entanglement nor composite systems \cite{R20, R21}.

A common theoretical technique to prove KS theorem is called parity proof. A set of \emph{R} rays (rank-1 projectors) and \emph{B} bases will said to provide a parity proof of the KS theorem if (a) \emph{B} is odd, and (b) each of the \emph{R} rays occurs an even number of times among the \emph{B} bases \cite{R10}. The set of rays and bases that satisfies these two conditions is called the KS set. The 13 rays used in \cite{R22} does not describe a KS set and an inequality instead of parity method is adopted to prove the KS theorem. Projectors in KS sets need not to be of rank-1 or all be of the same rank \cite{R10}. Kernaghan and Peres \cite{R9} provided a KS set in $\Re^8$ with 11 bases formed by a mixture of 16 rank-1 and 14 rank-2 projectors, which is labeled as 30-11, where the first figure is due to the sum of both rank-1 and rank-2 projectors. We put forward in this Letter a systematic way of constructing KS sets of this type. In the following, we firstly introduce the common properties of the KS sets of the type 30-11, followed by illustrating the simple steps of construction. We then provide a state-independent and a probabilistic state-specific proofs of KS theorem. We will end this Letter by a summary.

Since 1995, there is only one KS set in  $\Re^8$ with 11 bases formed by a mixture of 16 rank-1 and 14 rank-2 projectors \cite{R9}, see Table \ref{T1}. The first column in Table \ref{T1} gives the indices of 11 bases and the other four columns label the rank-1 and rank-2 projectors. Note that eigenvector (or ray) is represented mathematically by a column vector and rank-1 projector is obtained by taking outer product of the eigenvector. Hereafter, ray and rank-1 projector can be used interchangeably. Table 1 in \cite{R9} lists the 40 rays given by Kernaghan and Peres. The 40 rays are labeled as $R_i$, with \mbox{$i=$1, 2, 3, \textellipsis, 40}, and the corresponding rank-1 projectors would be labeled as $P_i$. On the other hand, rank-2 projectors are obtained by taking outer product for a linear combination of two rays. Table 2 of \cite{R10} lists explicitly all the 25 bases that can be formed by the 40 rays. The KS set given in Table \ref{T1} requires only 11 out of these 25 bases and only 36 rays are involved. Each of the projectors in Table \ref{T1} occurs twice among the 11 bases. In Table \ref{T1}, the parentheses $(i,j)$ refer to the rank-2 projectors that formed by rank-1 projectors $P_i$ and $P_j$. Due to the mixture of rank-1 and rank-2 projectors, there are 3 different basis sizes, i.e.,  1 basis of size 4 that formed by 4 rank-2 projectors, 4 bases of size 5 that formed by 3 rank-2 projectors and 2 rank-1 projectors and 6 bases of size 6 that formed by 2 rank-2 projectors and 4 rank-1 projectors. All these information is reflected via the symbol $16_2\mathbf{14_2}\mbox{-}1_44_56_6$ \cite{R10}, where the number of rank-2 projectors is typed in boldface, the first two subscripts indicate the multiplicities of projectors and the last three subscripts indicate the basis sizes.

All the rank-2 projectors in Table \ref{T1} is formed by 2 rank-1 projectors, thus it can easily be seen that the KS set shown in Table \ref{T1} is obtained from KS set of the type $28_28_4\mbox{-}11_8$ (refer to Table \ref{T2} which will be produced later), which contains only rank-1 projectors. As there are 320 KS sets of the type  $28_28_4\mbox{-}11_8$ \cite{R10}, it is natural to ask if there is any systematic way to transform them into KS sets of the type $16_2\mathbf{14_2}\mbox{-}1_44_56_6$. The answer is affirmative, as would be illustrated in the following.
\begin{table}[!h]
\caption{Kochen-Specker (KS) set of the type $16_2\mathbf{14_2}\mbox{-}1_44_56_6$ given by Kernaghan and Peres. Rank-2 projectors are indicated by parentheses.} \label{T1}
\begin{center}
\begin{tabular}{|c|c|c|c|c|}
  \hline
  5 & (\emph{33}, 35) & (\emph{34}, 40) & (\emph{36}, 37) & (\emph{38}, 39) \\ \hline
  10 & (\emph{34}, \emph{36}) & (\emph{33}, 35) & (\emph{8}, 2)  & 3, 5 \\ \hline
  22 & (\emph{33}, \emph{38}) & (\emph{34}, 40) & (\emph{18}, 19) & 21, 24 \\ \hline
  24 & (\emph{33}, \emph{38}) & (\emph{36}, 37) & (\emph{25}, 30) & 28, 31 \\ \hline
  16 & (\emph{34}, \emph{36}) & (\emph{38}, 39) & (\emph{12}, 9) & 14, 15 \\ \hline
  11 & (\emph{8}, 2) & (\emph{25}, 27) & 4, 6 & 26, 28 \\ \hline
  12 & (\emph{18}, 19) & (\emph{8}, 7) & 3, 4 & 17, 20 \\ \hline
  13 & (\emph{12}, 9) & (\emph{8}, 7) & 5, 6 & 10, 11 \\ \hline
  18 & (\emph{25}, 30) & (\emph{12}, 16) & 10, 14 & 26, 29 \\ \hline
  19 & (\emph{12}, 16) & (\emph{18}, 22) & 11, 15 & 17, 21 \\ \hline
  23 & (\emph{18}, 22) & (\emph{25}, 27) & 20, 24 & 29, 31 \\
  \hline
\end{tabular}
\end{center}
\end{table}
The first five bases in Table 2 of \cite{R10} are called pure bases ($PB$) and the rest are called hybrid bases ($HB$). All of the KS sets of the type $28_28_4\mbox{-}11_8$ contain 1 $PB$ and 10 $HB$s, and we indicate the former as $PB_i$, where the subscript refers to the index of that particular pure basis. We would label four of the rays from $PB_i$ as $\Gamma^i= \{ \alpha, \beta, \gamma, \delta \}$, if its subsets $\Gamma^i_a=\{ \alpha, \beta, \gamma \}$, $\Gamma^i_b=\{ \alpha, \beta, \delta \}$, $\Gamma^i_c=\{ \alpha, \gamma, \delta \}$ and $\Gamma^i_d=\{ \beta, \gamma, \delta \}$ appear in four different $HB$s of the KS set. The corresponding 4 $HB$s would then be labeled as $HB_a$, $HB_b$, $HB_c$ and $HB_d$, respectively. The remaining 4 rays of $PB_i$ form the set of $\neg \Gamma^i$. As indicated by $28_28_4\mbox{-}11_8$, there are 8 rays that each occurs 4 times, while four of them are contributed by $\Gamma^i$, the remaining 4 rays can be easily found by inspection and we would label them as $\Delta= \{ \epsilon, \zeta, \eta, \theta \}$. All the rays that each occurs 4 times in the KS set are italic typed in Table \ref{T1} (Table \ref{T2} as well).

\begin{figure}[!h]
\centering
  \includegraphics[width=2.4in]{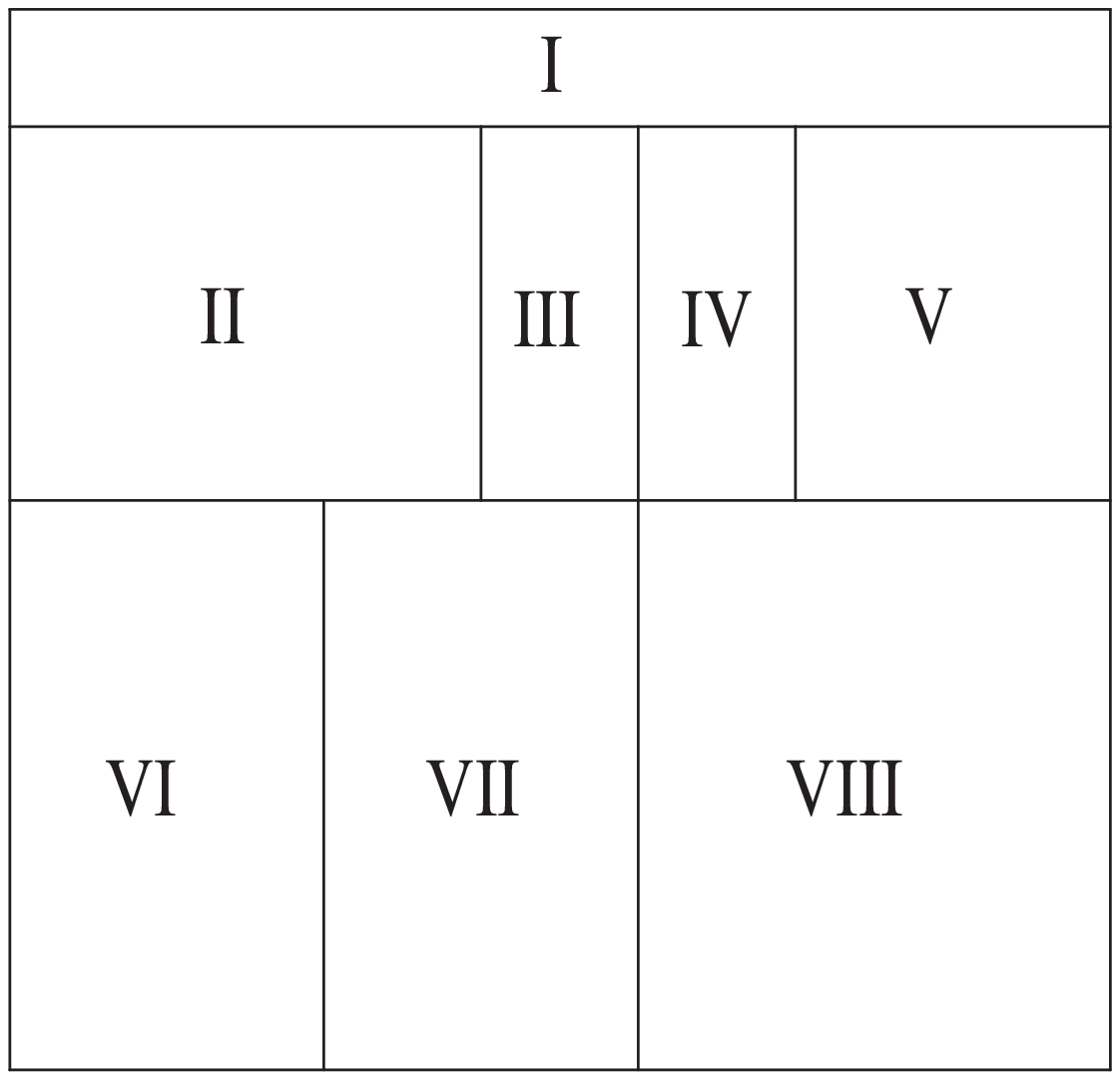}\\
  \caption{Partition off Table \ref{T2} results in eight divisions. The first column in Table \ref{T2} has been removed. Note that these divisions are commonly exist for all the KS sets of the type $28_28_4\mbox{-}11_8$} \label{F1}
\end{figure}

Given a specific KS set of the type $28_28_4\mbox{-}11_8$ with its bases listed in no particular order, we may rearrange the bases in such a way that $PB_i$ is taken as first basis, while $HB_a$, $HB_b$, $HB_c$ and $HB_d$ are placed as second, third, fourth and fifth basis, respectively. By taking the aforementioned $28_28_4\mbox{-}11_8$ KS set \cite{R9} as example, we therefore have $\Gamma^5= \{R33, R34, R36, R38 \}$, $\Gamma^5_{10}= \{R33, R34, R36 \}$, $\Gamma^5_{22}= \{R33, R34, R38 \}$, $\Gamma^5_{24}= \{R33, R36, R38 \}$ and $\Gamma^5_{16}= \{R34, R36, R38 \}$, where the subscripts $a$, $b$, $c$ and $d$ in this case are the indices for bases 10, 22, 24, and 16, respectively.  The remaining 6 bases can be listed arbitrarily. Table \ref{T2} shows the result obtained due to this rearrangement of bases. Meanwhile, we have $\neg \Gamma^5 = \{R35, R37, R39, R40 \}$ and $\Delta = \{R8, R12, R18, R25 \}$.
\begin{table}[!h]
\caption{Example for the KS set of the type $28_28_4\mbox{-}11_8$. The first column indicates the indices of bases. The 11 bases are rearranged in such a way that $PB_i$ is taken as first basis, followed by $HB_a$, $HB_b$, $HB_c$ and $HB_d$. In this case, $i=5$, $a=10$, $b=22$, $c=24$ and $d=16$. } \label{T2}
\begin{center}
\begin{tabular}{|c|c|c|c|c|c|c|c|c|}
   \hline
   5 & \emph{33} & \emph{34} & 35 & \emph{36} & 37 & \emph{38} & 39 & 40 \\ \hline
   10 & \emph{33} & \emph{34} & \emph{36} & 35 & \emph{8} & 2 & 3 & 5 \\ \hline
   22 & \emph{33} & \emph{34} & \emph{38} & 40 & \emph{18} & 19 & 21 & 24 \\ \hline
   24 & \emph{33} & \emph{36} & \emph{38} & 37 & \emph{25} & 28 & 30 & 31 \\ \hline
   16 & \emph{34} & \emph{36} & \emph{38} & 39 & \emph{12} & 9 & 14 & 15 \\ \hline
   11 & \emph{8} & \emph{25} & 2 & 28 & 4 & 6 & 26 & 27 \\ \hline
   12 & \emph{8} & \emph{18} & 3 & 19 & 4 & 7 & 17 & 20 \\ \hline
   13 & \emph{8} & \emph{12} & 5 & 9 & 6 & 7 & 10 & 11 \\ \hline
   18 & \emph{12} & \emph{25} & 14 & 30 & 10 & 16 & 26 & 29 \\ \hline
   19 & \emph{12} & \emph{18} & 15 & 21 & 11 & 16 & 17 & 22 \\ \hline
   23 & \emph{18} & \emph{25} & 24 & 31 & 20 & 22 & 27 & 29 \\
   \hline
 \end{tabular}
 \end{center}
\end{table}
We partition off Table \ref{T2} into 8 divisions and label them as $I, II, \textellipsis, VIII$, respectively, as shown in Figure \ref{F1}. Be cautious that in this partition, first column of Table \ref{T2} that indicates the indices of bases is excluded and therefore does not shown in Figure \ref{F1}. The rays of $\Gamma^5_{10}$, $\Gamma^5_{22}$, $\Gamma^5_{24}$ and $\Gamma^5_{16}$ are located at $II$, the rays of $\neg \Gamma^5$ are located at $III$ and the rays of $\Delta$ are located at IV. Every ray in IV will repeat three times in VI and every ray in $V$ repeats its second time in $VII$. For the remaining 12 rays, each occurs twice in VIII.
We name the rank-2 projectors formed by 2 rank-1 projectors, both of which repeat 4 times, as type-A projectors. On the other hand, we name the rank-2 projectors, formed by 1 rank-1 projector that repeats 4 times and another rank-1 projector that repeats twice, as type-B projectors. There are 4 type-B projectors in $I$ and 4 type-A projectors in $II$. There are 4 type-B $II\mbox{-}III$ projectors, where $II\mbox{-}III$ means that 1 rank-1 projector from $II$ and another rank-1 projector from $III$ are chosen in order to form the rank-2 projectors. There are 4 type-B $IV\mbox{-}V$ projectors, 4 type-B $VI\mbox{-}VII$ projectors and 8 type-B $VI\mbox{-}VIII$ projectors. Such constructed 28 rank-2 projectors all are shown in Table \ref{T1} and they in fact consist of 14 different rank-2 projectors. It can be easily seen from Table \ref{T1} that there are 2 type-A projectors that each repeats twice in $II$. Each of the 4 type-B $II\mbox{-}III$ projectors repeats its second time in $I$; The 4 type-B $IV\mbox{-}V$ projectors are exactly the same as 4 type-B $VI\mbox{-}VII$ projectors. Finally, there are 4 type-B $VI\mbox{-}VIII$ projectors that each of them occurs twice.

The number of ways of picking the rank-1 projectors to form rank-2 projectors determine the number of KS sets of the type $16_2\mathbf{14_2}\mbox{-}1_44_56_6$ that can be transformed from a given specific KS set of the type $28_28_4\mbox{-}11_8$. There are 7 different ways of forming type-B projectors in $I$ and the construction of type-A projectors in $II$ is completely confined by the former. On the other hand, the formation of type-B $VI\mbox{-}VII$ and $VI\mbox{-}VIII$ projectors are completely confined by the ways of forming type-B $IV\mbox{-}V$ projectors and there are 81 ways for the latter. Therefore, there are 567 KS sets of the type $16_2\mathbf{14_2}\mbox{-}1_44_56_6$ that can be generated from each of the KS sets of the type $28_28_4\mbox{-}11_8$.

We now provide a state-independent proof using KS set generated above. For every basis of the KS set listed in Table \ref{T1}, $\sum P_i =I$, where $I$ is the eight-dimensional identity matrix and $P_i$ are either rank-1 or rank-2 projectors. As QM follows sum rule, it means that $\sum v(P_i)=v(I)$, where $v(\cdot)$ is the value function that assigns values 1 or 0 to the projectors. The $v(I)$ always gives value 1, this implies that there is one and only one projector on the left hand side of the equality sign that would be assigned value 1 but the rest would be assigned value 0. There are 11 bases shown in Table \ref{T1}, each of the rank-1 and rank-2 projectors among them occurs twice. For the same projectors, they would be assigned the same values (0 or 1) even though they are in different bases, if the assumptions of NCHV theories are valid, i.e., the values of the measurements pre-exist, and the results of measurements are noncontextual. However, the noncontextual values assignment will lead to a contradiction, because the sum of the 11 $\sum v(P_i)$ on the left of the equality signs will give an even number (since each of the $v(P_i)$ occurs twice), but the 11 $v(I)$ on the right of the equality signs will give an odd number, i.e., 11. This provides a state-independent parity proof of the KS theorem.

Note that, by recovering the rank-2 projector $P_{25, 27}$ (refer to Table \ref{T1}) to rank-1 projectors $P_{25}$ and $P_{27}$, the KS set of the type $16_2\mathbf{14_2}\mbox{-}1_44_56_6$ shown in Table \ref{T1} will be changed to the KS set of the type $18_2\mathbf{13_2}\mbox{-}1_44_54_62_7$. Using the latter KS set, a probabilistic state-dependent proof of KS theorem \cite{R6} exists as follows. Firstly, suppose  that a system of state $\vert R_{33} \rangle$ is preselected, then $v(P_{33})=1$; Secondly, in a post-selection, assume that we find the system in the state $\vert R_{12} \rangle$, then $v(P_{12})=1$. This is possible because $\langle R_{12} \vert R_{33} \rangle \neq 0$. All the rank-1 and rank-2 projectors that are orthogonal to either $P_{33}$ or $P_{12}$ would be assigned a value 0. Consequently, all the rank-2 projectors are being removed and only 3 rank-1 projectors remain in the following equations,
\begin{eqnarray} \label{eq1}
v(\hat{P}_{27}) + v(\hat{P}_4)=v(\hat{I}), \notag \\
v(\hat{P}_4) + v(\hat{P}_{20}) =v(\hat{I}), \notag \\
v(\hat{P}_{27}) + v(\hat{P}_{20})= v(\hat{I}).
\end{eqnarray}
Again, as each of the 3 rank-1 projectors occurs twice and there are only 3 $v(I)$ in (\ref{eq1}), it is impossible to have a consistent way of values assignment (0 or 1) for these projectors. In this demonstration of the KS theorem proof, 20 rank-1 projectors are used, i.e., 1 contributed by $P_{33}$, 9 contributed by rank-1 projectors that are perpendicular to $P_{33}$ ($P_3$, $P_5$, $P_{10}$, $P_{11}$, $P_{21}$, $P_{24}$, $P_{25}$, $P_{28}$ and $P_{31}$), 1 contributed by $P_{12}$, 6 contributed by rank-1 projectors that are perpendicular to $P_{12}$ ($P_6$, $P_{14}$, $P_{15}$, $P_{17}$, $P_{26}$, $P_{29}$) and 3 contributed by rank-1 projectors that occur in (\ref{eq1}).

The violation of KS theorem means that QM cannot be regarded as a noncontextual hidden variable theory. In 1995, Kernaghan and Peres \cite{R9} proved KS theorem by giving a KS set with a mixture of 16 rank-1 projectors and 14 rank-2 projectors in eight-dimensional state space, which is represented by $16_2\mathbf{14_2}\mbox{-}1_44_56_6$ \cite{R10}. The KS set may be constructed based on intuition and there have not been any other copies proposed since then. We give a systematic way of construction and show that the number of the KS sets of the type $16_2\mathbf{14_2}\mbox{-}1_44_56_6$ that can be produced is indeed large. In addition to the standard method of state-independent proof of the KS theorem, we also provide a version of probabilistic  state-dependent proof of the KS theorem that underscores values assignment in such a way to remove all the rank-2 projectors from the KS set. This method allows us to use only 20 rank-1 projectors in the proof.

This work is supported by the Ministry of Higher Education of Malaysia under the Fundamental Research Grant Scheme, FRGS/1/2011/ST/UNIM/03/1.


%




\begin{thebibliography}{99}
\bibitem{R1} Peres A 1993 \emph{Quantum Theory: Concepts and Methods} (Kluwer, Dordrecht) p196
\bibitem{R2} Kochen S and Specker E P 1967 \emph{J.\ Math.\ Mech.\ } \textbf{17} 59
\bibitem{R3} Peres A 1991 \emph{J.\ Phys.\ A: Math.\ Gen.\ } \textbf{24} L175
\bibitem{R4} Conway J H and Kochen S, reported by Peres A 1993 in \emph{Quantum Theory: Concepts and Method} (Kluwer, Dordrecht) p114
\bibitem{R5} Kernaghan M 1994 \emph{J.\ Phys.\ A: Math. Gen.\ } \textbf{27} L829
\bibitem{R6} Cabello A, Estebaranz J M and Garc\'{i}a-Alcaine G 1996 \emph{Phys.\ Lett.\ A} \textbf{212} 183
\bibitem{R7} Maegell M, Aravind P K, Megill N D and Pavi$\check{\mbox{c}}$i$\acute{\mbox{c}}$ M,  2011 \emph{Found.\ Phys.\ } \textbf{41} 883
\bibitem{R8} Cabello A and Garc\'{i}a-Alcaine G 2005 \emph{Phys.\ Lett.\ A} \textbf{339} 425
\bibitem{R9} Kernaghan M and Peres A 1995 \emph{Phys.\ Lett.\ A} \textbf{198} 1
\bibitem{R10} Waegell M and Aravind P K 2012 \emph{J.\ Phys.\ A: Math.\ Theor.\ } \textbf{45} 405301
\bibitem{R11} Planat M 2012 \emph{EPJ Plus} \textbf{127} 86
\bibitem{R12} Michler M, Weinfurter H and Zukowski M 2000 \emph{Phys\ Rev.\ Lett.\ } \textbf{84} 5457
\bibitem{R13} Huang Y F, Li C F, Zhang Y S, Pan J W, and Guo G C 2003 \emph{Phys.\ Rev.\ Lett.\ } \textbf{90} 250401
\bibitem{R14} Yang T, Zhang Q, Yin J, Zhao Z, Zukowski M, Chen Z B and Pan J W 2005 \emph{Phys.\ Rev.\ Lett.\ } \textbf{95} 240406
\bibitem{R15} Amselem E, Radmark M, Bourennane M and Cabello A 2009 \emph{Phys.\ Rev.\ Lett.\ } \textbf{103} 160405
\bibitem{R16} Hasegawa Y, Loidl R, Badurek G, Baron M, and Rauch H 2006 \emph{Phys.\ Rev.\ Lett.\ } \textbf{97} 230401
\bibitem{R17} Bartosik H, Klepp J, Schmitzer C, Sponar S, Cabello A, Rauch H and Hassegawa Y 2009 \emph{Phys.\ Rev.\ Lett.\ } \textbf{103} 040403
\bibitem{R18} Kirchmair G, Z$\ddot{\mbox{a}}$hringer F, Gerritsma R, Kleinmann M, G$\ddot{\mbox{u}}$hne O, Cabello A, Blatt R and Roos C F 2009 \emph{Nature} \textbf{460} 494
\bibitem{R19} Wei L F, Maruyama K, Wang X B, You J Q and Nori F 2010 \emph{Phys. Rev.\ B} \textbf{81} 174513
\bibitem{R20} Liu B H, Huang Y F, Gong Y X, Sun F W, Zhang Y S, Li C F and Guo G C 2009 \emph{Phys.\ Rev.\ A} \textbf{80} 044101
\bibitem{R21} Simon C, $\dot{\mbox{Z}}$ukowski M, Weinfurter H and Zeilinger A 2000 \emph{Phys.\ Rev.\ Lett.\ } \textbf{85} 1783
\bibitem{R22} Yu S and Oh C H 2012 \emph{Phys.\ Rev.\ Lett.\ } \textbf{108} 030402
\end{thebibliography}
\end{document}